\documentclass[]{spie}  

\usepackage{amsmath,amsfonts,amssymb}
\usepackage{caption}
\usepackage{graphicx}
\usepackage[colorlinks=true, allcolors=blue]{hyperref}

\usepackage{ulem}

\title{SCExAO, a testbed for developing high-contrast imaging technologies for ELTs}

\author[a]{Kyohoon Ahn}
\author[a,b,c,d]{Olivier Guyon}
\author[a]{Julien Lozi}
\author[a,d,e]{S\'ebastien Vievard}
\author[a]{Vincent Deo}
\author[a,e,f]{Nour Skaf}
\author[g]{Ruslan Belikov}
\author[h]{Steven P. Bos}
\author[i]{Michael Bottom}
\author[a,g]{Thayne Currie}
\author[j]{Richard Frazin}
\author[b]{Kyle V. Gorkom}
\author[k]{Tyler D. Groff}
\author[b]{Sebastiaan Y. Haffert}
\author[l]{Nemanja Jovanovic}
\author[m]{Hajime Kawahara}
\author[d,n]{Takayuki Kotani}
\author[b]{Jared R. Males}
\author[o]{Frantz Martinache}
\author[p]{Benjamin A. Mazin}
\author[h]{Kelsey Miller}
\author[q]{Barnaby Norris}
\author[b]{Alexander Rodack}
\author[q]{Alison Wong}

\affil[a]{Subaru Telescope, National Astronomical Observatory of Japan, National Institutes of Natural Sciences (NINS), 650 North A`oh\={o}k\={u} Place, Hilo, HI 96720, United States}
\affil[b]{Steward Observatory, University of Arizona, Tucson, AZ 87521, United States}
\affil[c]{College of Optical Sciences, University of Arizona, Tucson, AZ 87521, United States}
\affil[d]{Astrobiology Center of NINS, 2 Chome-21-1, Osawa, Mitaka, Tokyo, 181-8588, Japan}
\affil[e]{Observatoire de Paris, LESIA, 5 Place Jules Janssen, 92190 Meudon, France}
\affil[f]{Department of Physics and Astronomy, University College London, London, United Kingdom}
\affil[g]{NASA Ames Research Center, Moffett Blvd, Mountain View, CA 94035, United States}
\affil[h]{Leiden Observatory, Leiden University, Niels Bohrweg 2, 2333 CA, Leiden, The Netherlands}
\affil[i]{Institute for Astronomy, University of Hawai'i, 640 North A`oh\={o}k\={u} Place, Hilo, HI 96720, United States}
\affil[j]{Department of Climate and Space Sciences and Engineering, University of Michigan, Ann Arbor, MI 48109, United States}
\affil[k]{Goddard Space Flight Center, 8800 Greenbelt Rd, Greenbelt, MD 20771, United States}
\affil[l]{California Institute of Technology, 1200 E California Blvd, Pasadena, CA 91125, United States}
\affil[m]{University of Tokyo, 7 Chome-3-1 Hongo, Bunkyo City, Tokyo 133-8654, Japan}
\affil[n]{National Astronomical Observatory of Japan, NINS, 2 Chome-21-1, Osawa, Mitaka, Tokyo, 181-0015, Japan}
\affil[o]{Laboratoire Lagrange, Universit\'e C\^{o}te d'Azur, Observatoire de la C\^{o}te d'Azur, CNRS, Parc Valrose, B\^{a}t. H. FIZEAU, 06108 Nice, France}
\affil[p]{University of California Santa Barbara, Santa Barbara, CA 93106, United States}
\affil[q]{Sydney Institute for Astronomy, Institute for Photonics and Optical Science, School of Physics, University of Sydney, NSW 2006, Australia}

\authorinfo{Further author information: (Send correspondence to K.A.)\\K.A.: E-mail: kyohoon@naoj.org, Telephone: 1 808 202 1095}

\begin{document} 
\maketitle

\begin{abstract}
To directly detect exoplanets and protoplanetary disks, the development of high accuracy wavefront sensing and control (WFS\&C) technologies is essential, especially for ground-based Extremely Large Telescopes (ELTs). The Subaru Coronagraphic Extreme Adaptive Optics (SCExAO) instrument is a high-contrast imaging platform to discover and characterize exoplanets and protoplanetary disks. It also serves as a testbed to validate and deploy new concepts or algorithms for high-contrast imaging approaches for ELTs, using the latest hardware and software technologies on an 8-meter class telescope. SCExAO is a multi-band instrument, using light from 600 to 2500 nm, and delivering a high Strehl ratio ($>$80\% in median seeing in H-band) downstream of a low-order correction provided by the facility AO188. Science observations are performed with coronagraphs, an integral field spectrograph, or single aperture interferometers. The SCExAO project continuously reaches out to the community for development and upgrades. Existing operating testbeds such as the SCExAO are also unique opportunities to test and deploy the new technologies for future ELTs. We present and show a live demonstration of the SCExAO capabilities (Real-time predictive AO control, Focal plane WFS\&C, etc) as a host testbed for the remote collaborators to test and deploy the new WFS\&C concepts or algorithms. We also present several high-contrast imaging technologies that are under development or that have already been demonstrated on-sky.
\end{abstract}


\section{INTRODUCTION}
\label{sec:intro}  
Since 2014, High-Contrast Imaging (HCI) capabilities have been deployed on large ground-based telescops with several Extreme Adaptive Optics (ExAO) systems such as the Gemini Planet Imager (GPI)\cite{macintosh2014first} and the Spectro-Polarimetric High-contrast Exoplanet REsearch instrument (SPHERE)\cite{beuzit2008sphere}. These systems all share a similar architecture: they employ a high-order wavefront sensor (WFS) and a deformable mirror (DM) to correct for atmospheric turbulence enabling high Strehl ratio in the near-infrared (NIR) ($\approx$90\%), while a coronagraph is used to suppress on-axis starlight downstream. The main goal of these systems is the direct detection of exoplanets and planet-forming disks \cite{Macintosh2015,Boccaletti2015,Chauvin2017,Currie2015,Keppler2018,Haffert2019,Bohn2020}. Moreover, these systems are crucial to test the necessary technologies that future HCI system for future Extremely Large Telescopes (ELTs)\cite{sanders2013thirty, johns2006giant, gilmozzi2007european} will require. The Subaru Coronagraphic Extreme Adaptive Optics (SCExAO)\cite{jovanovic2015subaru} has also evolved into a HCI system for past 12 years. Likewise, it can be used as a testbed to test and deploy new concepts or algorithms for HCI approaches for future ELTs, using the latest hardware and software technologies on an 8-meter class telescope.
In this paper, we present the current status and overview of the SCExAO, especially the main components for the HCI system and science modules that can be used for competitive science. Also, we present the current collaborations including hardware and algorithms. Finally, we introduce the detail of the major upgrades planned for SCExAO and AO188, that will bring the instrument closer to a system-level demonstrator for future ELTs.

\section{Overview of SCE\lowercase{x}AO}
\label{sec:overview}  
\begin{figure}[h]
    \centering
    \includegraphics{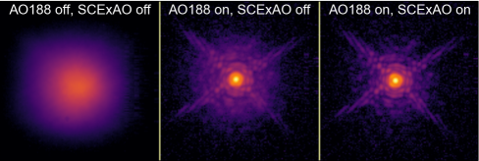}
    \vspace*{0.3cm}
    \caption{On-sky PSF in seeing limited mode (left), after the first stage of correction by AO188 (middle), after SCExAO correction (right). With SCExAO correction, the PSF is more stable and the speckle halo surrounding the core is fainter.}
    \label{fig:on-sky_PSF}
\end{figure}

SCExAO is a HCI system installed on the infrared Nasmyth platform at the Subaru telescope behind the facility adaptive optics, AO188\cite{minowa2010performance} that performs a first level of atmospheric turbulence correction. Typical Strehl ratios are 20 to 40\% after the correction using the AO188. In addition, the SCExAO provides the second level of correction using a 2,000-actuator DM and a visible Pyramid WFS (PyWFS) and allowing to obtain a high Strehl ration($\approx$ 90\% with good seeing in NIR)\cite{jovanovic2015subaru,lozi2019visible}. Figure \ref{fig:on-sky_PSF} shows the effect of the first level correction by AO188, and the ExAO correction by SCExAO. With the SCExAO correction, the speckle halo around the point spread function (PSF) is more stable, and at a lower level, allowing for better sensitivity when looking for companions and disk near host stars.

\begin{figure}[h]
    \centering
    \includegraphics[width=12cm]{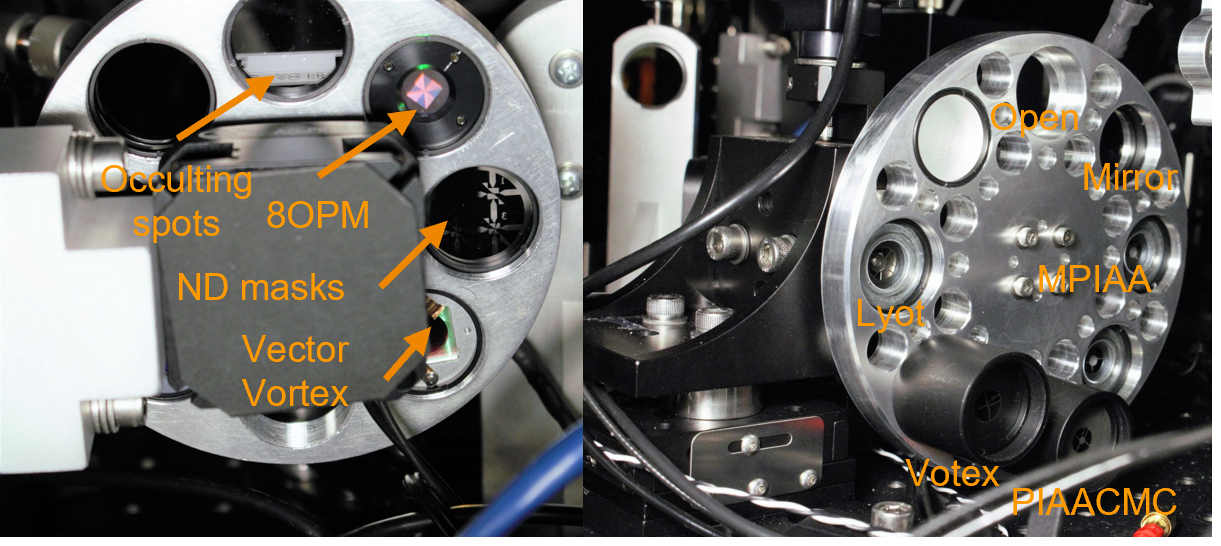}
    \vspace*{0.3cm}
    \caption{Focal plane masks (left) and Lyot stops (right) are installed in filter wheels}
    \label{fig:FPM_Lyot}
\end{figure}

A suite of coronagraphs in the infrared path suppresses starlight. As shown in Fig. \ref{fig:FPM_Lyot}, there are nine occulting spots spanning a range of Inner Working Angle (IWA).Coronagraph options include classical Lyot coronagraph, and small IWA coronagraphs, such as the vector vortex coronagraph, the 8-octant phase mask (8OPM) coronagraph\cite{nishikawa2020combination}, and the Phase-Induced Amplitude Apodization Complex Mask Coronagraph (PIAACMC)\cite{knight2018phase}. These coronagraphs offer the lowest IWA and high throughput but are more sensitive to wavefront error. On the other hand, there are pupil masks changing the diffraction pattern for HCI, such as the shaped pupil\cite{currie2018laboratory} and vector Apodizing Phase Plate (vAPP)\cite{bos2018fully}. These pupil masks offer a larger IWA and lower throughput but are less sensitive to residual wavefront error. Hence, the coronagraphs available are designed to span a large range of residual wavefront error and should be chosen accordingly. The main one used currently is the classical Lyot coronagraph, which is a dot masking the star in the focal plane used in combination with a pupil plane Lyot stop.

\begin{figure}[h]
    \centering
    \includegraphics[width=12cm]{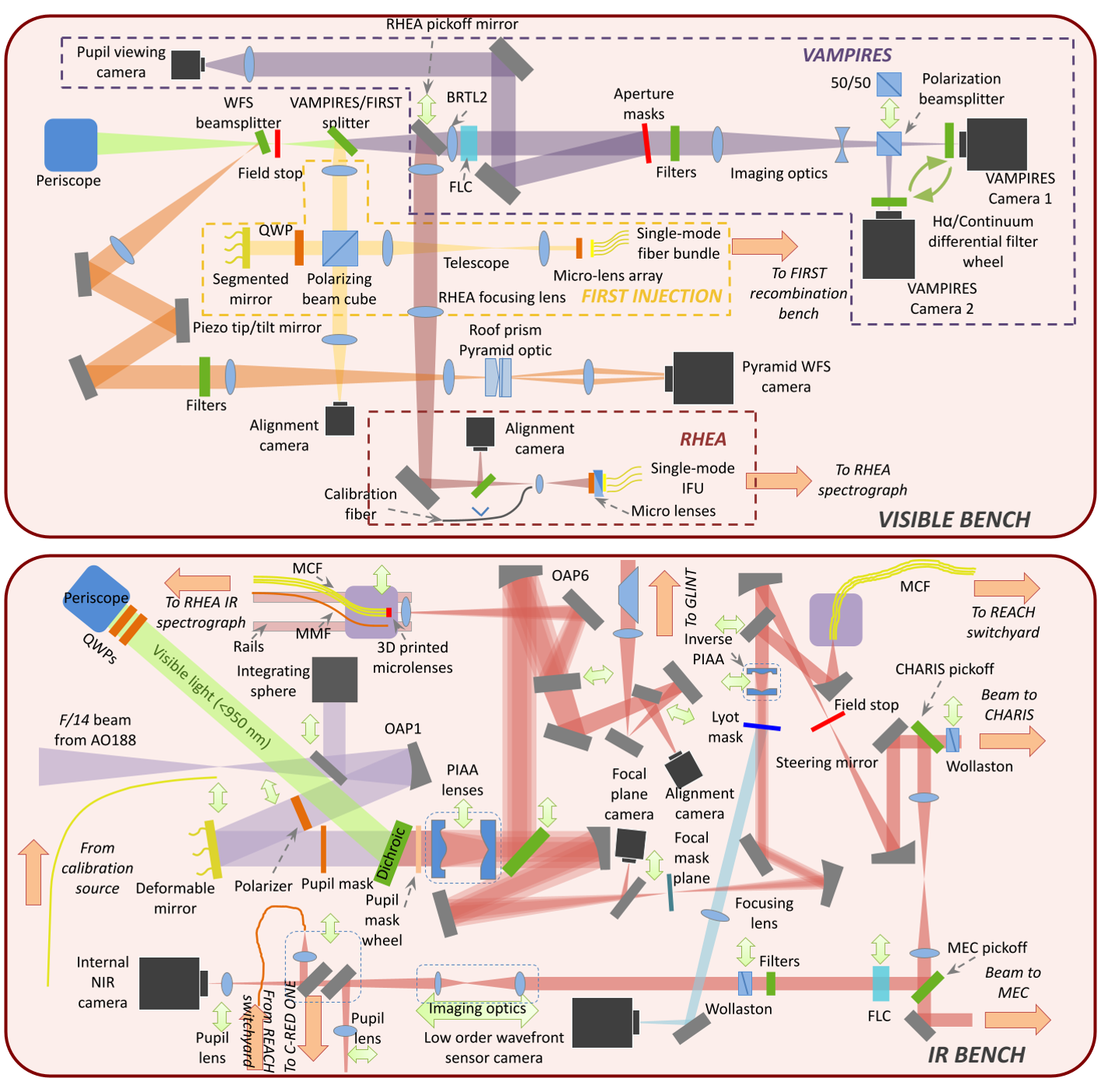}
    \vspace*{0.3cm}
    \caption{Schematic diagram of the SCExAO instrument. Top box: layout of the visible optical bench whichis mounted on top of the IR bench. Bottom box: IR bench layout. Dual head green arrows indicate that a given optics can be translated in/out of or along the beam. Orange arrows indicate light entering or leaving the designated bench at that location.}
    \label{fig:IR_and_Visible}
\end{figure}

As shown in Fig. \ref{fig:IR_and_Visible}, the instrument adopts a modular design and consists of separate visible ($<$940 nm) and infrared (IR, $>$940 nm) benches. This modular design allows concurrent operation of multiple science modules. The IR bench hosts the DM, coronagraphs, and multi-core fiber bundle for REACH (Rigorous Exoplanetary Atmosphere Characterization with High dispersion coronagraphy)\cite{kotani2020reach} module and sends light to the CHARIS (Coronagraphic High Angular Resolution Imaging Spectrograph)\cite{groff2013design} and GLINT (Guided-Light Interferometric Nulling Technology)\cite{martinod2020first} modules, while the visible bench hosts the PyWFS, VAMPIRES (Visible Aperture Masking Polarimetric Imager Resolved Exoplanetary Structures)\cite{norris2020diffraction}, and FIRST (Fibered Imager foR a Single Telescope)\cite{huby2013first} modules. These science modules can be fed with a suite of dichroics and beamsplitters. 

The main science is performed by Integral Field Spectrograph (IFS) CHARIS using J-, H-, and K-band (1.1 to 2.4 \textmu m). CHARIS has a field of view 2 by 2 arcsecond with two different modes. The low-resolution mode (R$\approx$20) uses broadband light between J- and K-bands while the high-resolution mode (R$\approx$70) performs single band spectroscopy at J-, H-, or K-band. Using CHARIS typically observes exoplanets, protoplanetary and debris disks and performs spectral characterization\cite{CurrieSPIE,Currie2018,Goebel2018,Currie2019,Currie2020,Uyama2020,Lawson2020,Steiger2021}.

\begin{figure}[h]
    \centering
    \includegraphics[width=15cm]{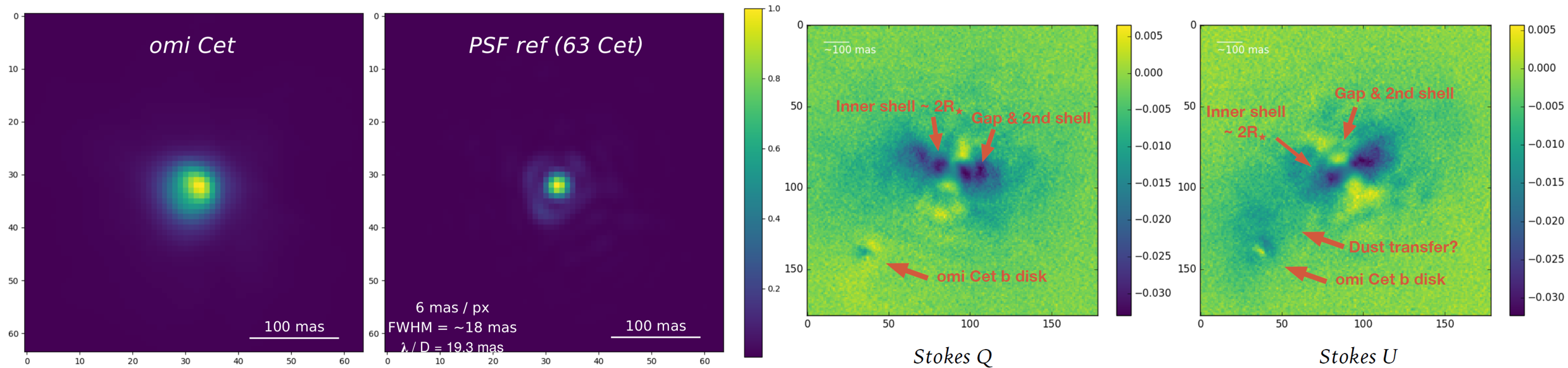}
    \vspace*{0.3cm}
    \caption{Examples of results obtained with the VAMPIRES: The post-AGB star Mira (\textit{o} Cet) is resolved by the VAMPIRES, compared to a reference star. In polarized light, shells of dust are visible around Mira A, while a disk is also visible around its companion Mira B.}
    \label{fig:vamp_result}
\end{figure}

In visible, the VAMPIRES module performs polarization differential imaging (PDI) and can be combined with aperture masking interferometry to reach higher spatial resolution. It operates from 600 to 800 nm and allow for sub-diffraction limited imaging of post AGB star Mira and disks. Figure \ref{fig:vamp_result} shows some results obtained with VAMPIRES on the post-AGB star Mira.

Another module that is now available for science is REACH. REACH connects the high-contrast imaging capability of the SCExAO to the a high-resolution spectrograph IRD (InfraRed Doppler)\cite{kotani2018infrared}, which has a resolution of 100,000 between Y- and H-band. The fiber used for the injection has multiple cores, allowing for a core to sample the light from a companion (exoplanet, brown dwarf) and other cores to sample only the residual light from the central star.

In addition, there are two new modules recently opened for science on the SCExAO. The first one is the Fast-IR PDI mode using a fast low-noise IR detector(C-RED one camera, which can operate at a few tens of kilohertz) synchronized to a polarization-switching Ferroelectric Liquid Crystal (FLC), similarly to the VAMPIRES. The second one, MEC (MKID Exoplanet Camera)\cite{walter2020mkid} is an innovative type of noiseless photon-counting energy resolving detector. MEC measures each photon's arrival time and energy, which means it can do low-resolution spectroscopy without a dispersive element. The detector's timing and energy resolution are leveraged by the SSD (Stochastic Speckle Discrimination) method\cite{meeker2018darkness}.

\section{Current Collaborations of SCE\lowercase{x}AO for HCI technologies}

\begin{figure}[h]
    \centering
    \includegraphics[width=15cm]{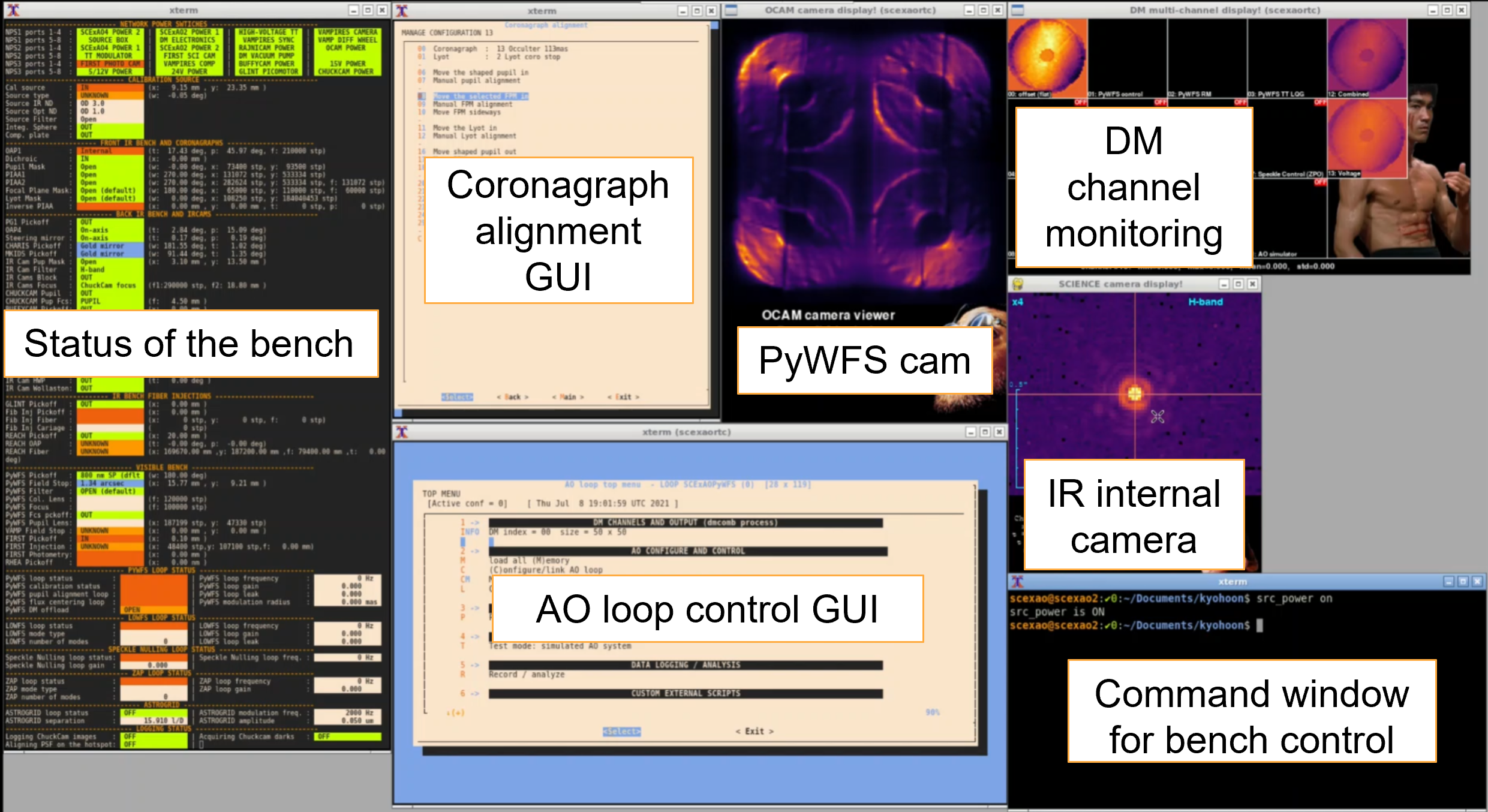}
    \vspace*{0.3cm}
    \caption{The main VNC window controlling the hardware of the instrument, with a status of the various motors, control of the coronagraphs, and display of the different cameras, DM, and AO loop control GUI.}
    \label{fig:VNC}
\end{figure}

Beside the classical HCI system, SCExAO can also be used as a testbed for developing HCI technologies. For that, we developed a platform that can be used day or night from anywhere in the world. The instrument is available almost 24/7 for collaborators to use ---minus observing nights and when the instrument is down for maintenance---. We usually provide VPN access and once that access is granted, remote users can control the instrument and test custom algorithms. Also, we provide some training on the different necessary procedures. The core team in Hilo, Hawaii is available to assist for testing, and can also help the hardware changes or alignments at the telescope. Figure \ref{fig:VNC} shows an examples of a VNC terminal used to control the hardware on the instrument. When the tests are successful in the laboratory conditions, they can be tested on-sky with about six to seven nights of engineering time allocated per year.

\subsection{Hardware Collaborations}
On the hardware side, we tested a variety of coronagraphs, which are mentioned in Section \ref{sec:overview}, such as the vector vortex coronagraph, the 8OPM, PIAACMC, the shaped pupil, and the vAPP. We are also investigating ways of improving the astrometric and photometric calibration of the science images, when the coronagraph is in place. While we currently use a temporally modulated speckle grid created by the deformable mirror, we showed that adding a spatial modulation increased the precision of the calibration\cite{sahoo2020precision}. Other techniques are also investigated, such as an incoherent speckle grid created by a polarizing phase plate\cite{bos2020vector}.

\subsection{Interferometric and Fiber injection modules}
We also developed a number of innovative interferometry and fiber injection modules that are used for science, but also as wavefront sensors:

\begin{itemize}
    \item \textbf{FIRST:} a visible interferometric module with spectroscopic capabilities (see S. Vievard et al. from this conference\cite{VievardSPIE}),
    \item \textbf{GLINT:} a NIR photonic nulling interferometer (see S. Vievard et al. from this conference\cite{VievardSPIE}),
    \item \textbf{NRM:} wavefront sensing using non-redundant aperture masking interferometry (see V. Deo et al. from this conference\cite{DeoSPIE}),
    \item An innovative fiber injected low-order wavefront sensor using a photonic lantern\cite{norris2020all}.
\end{itemize}

\subsection{Focal plane wavefront sensing and control}

Beside hardware collaborations, the most active collaborations on the instrument are related to focal plane wavefront sensing and control algorithms. For the past few years, we tested several algorithms aimed at measuring and correcting low-order aberrations, quasi-static non-common path aberration, or low-wind/island effect. Since SCExAO has only one DM, the commands from the other correction loops are applied by offsetting the reference of the ExAO loop.

\begin{figure}[h]
    \centering
    \includegraphics{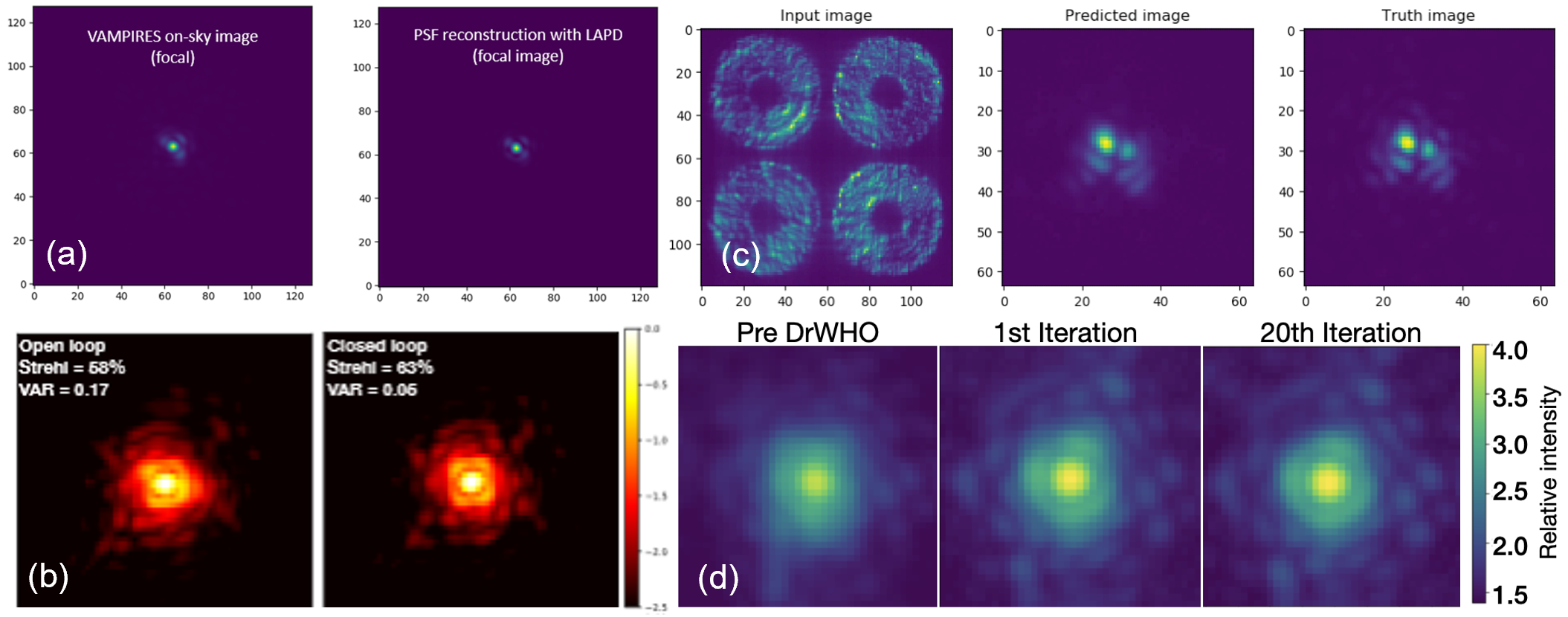}
    \vspace*{0.3cm}
    \caption{Examples of focal plane WFS/C algorithms for low-order aberrations, (a) LAPD\cite{vievard2020cophasing}, (b) Fast\&Furious\cite{bos2020sky}, (c) focal plane image reconstruction from PyWFS images using a NN\cite{wong2021predictive}, and (d) DrWHO. All images are from on-sky results.}
    \label{fig:Low_orders}
\end{figure}

The algorithms tested so far are (see example in Fig. \ref{fig:Low_orders}):
\begin{itemize}
    \item \textbf{Zernike Asymmetric Pupil (ZAP):} a phase retrieval algorithm using an asymmetric pupil mask\cite{n2018calibration},
    \item \textbf{Phase diversity algorithms:} Linearized Analytic phase diversity (LAPD)\cite{vievard2020cophasing}, Mono-plane phase diversity, Fast\&Furious\cite{bos2021first},
    \item PSF reconstruction from PyWFS images using a neural network (NN)\cite{wong2021predictive},
    \item \textbf{Direct reinforcement Wavefront Heuristic Optimization (DrWHO):} an algorithm based on reinforcement learning (see N. Skaf et al. from this conference\cite{Skaf2021DrWHOSPIE}).
\end{itemize}

\begin{figure}[h]
    \centering
    \includegraphics[width=12cm]{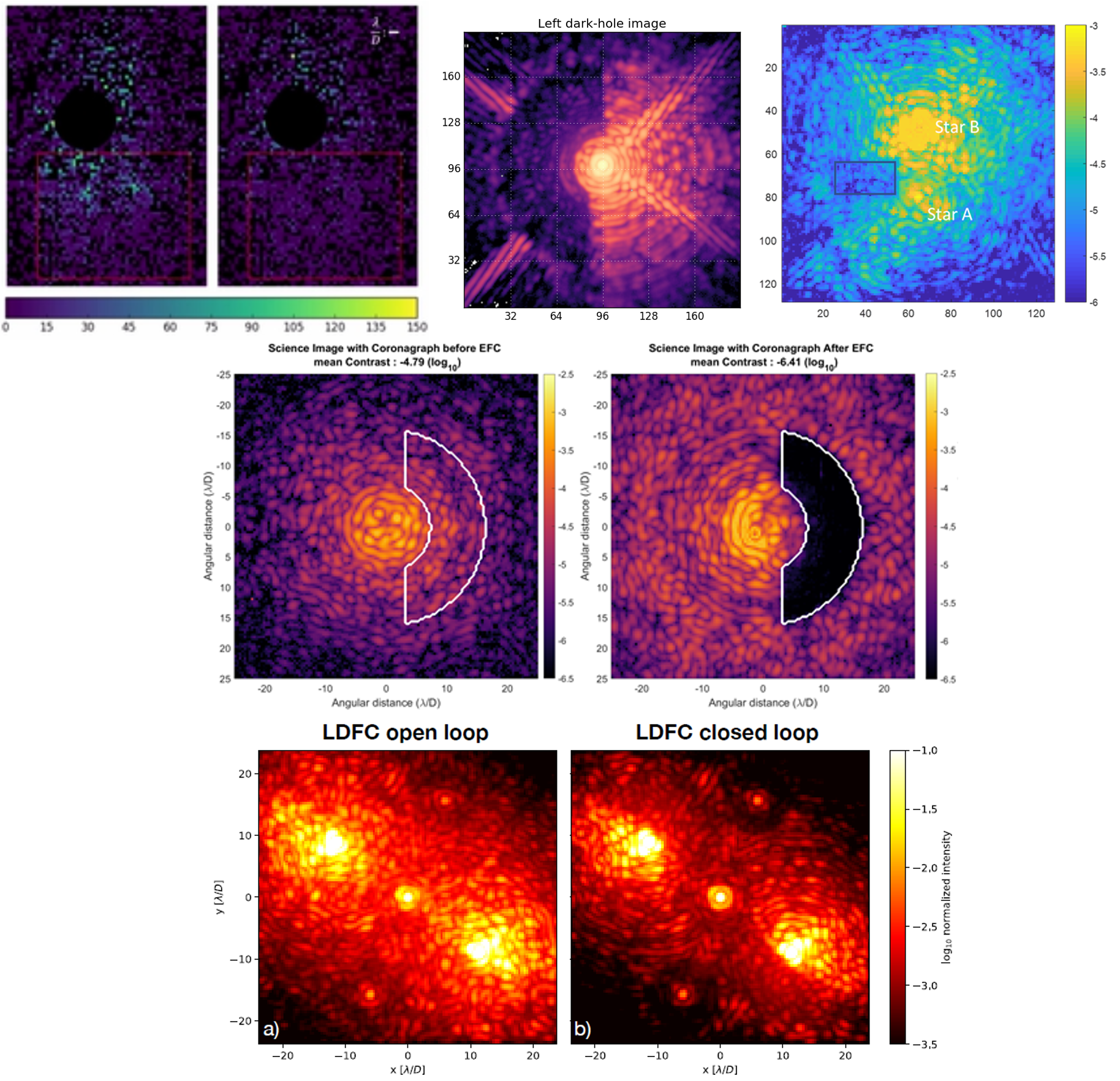}
    \vspace*{0.3cm}
    \caption{Examples of focal plane WFS/C algorithms for high-order aberrations, On-sky results of the speckle nulling with the MEC, a laboratory result of the speckle nulling on SCExAO, and a laboratory result of the MSWC (top row from left to right), a laboratory results of the EFC (middle row), and on-sky result of the dark hole stabilization using the LDFC (bottom row).}
    \label{fig:speckle_controls}
\end{figure}

In addition to the correction of low-order aberrations, we implemented several speckle control techniques such as speckle nulling\cite{martinache2014sky}, electric field conjugation (EFC), Linear Dark Field Control (LDFC)\cite{bos2021first}, necessary to reach deeper contrasts by correcting high-order aberrations unseen by the PyWFS. Their on-sky implementation was proven quite difficult, as they tend to require a very stable wavefront correction, and a good calibration of the PyWFS. For on-sky implementation, we have been using fast infrared cameras such as the C-RED1, the C-RED2, and the MEC. We also tested less conventional speckle nulling algorithm, such as the multi-star wavefront control (MSWC)\cite{belikov2016high}, a speckle nulling algorithm used for binary systems. Finally, we demonstrated the LDFC algorithm to stabilize the dark hole over longer periods of time, notably by measuring the disturbances using the opposite bright field. Figure \ref{fig:speckle_controls} shows the results of  the speckle nulling, the EFC, and the LDFC algorithm. Especially, the laboratory demonstration of the EFC is uploaded on the YouTube channel of the SCExAO team (see \url{https://youtu.be/ppQupxlmi8U}). These collaborations are helping solve new challenges that will face future high-contrast imaging systems, and therefore are essential for reaching the contrast necessary to detect earth-like planets with ELTs.

\section{Future upgrades}
\begin{figure}[h]
    \centering
    \includegraphics[width=15cm]{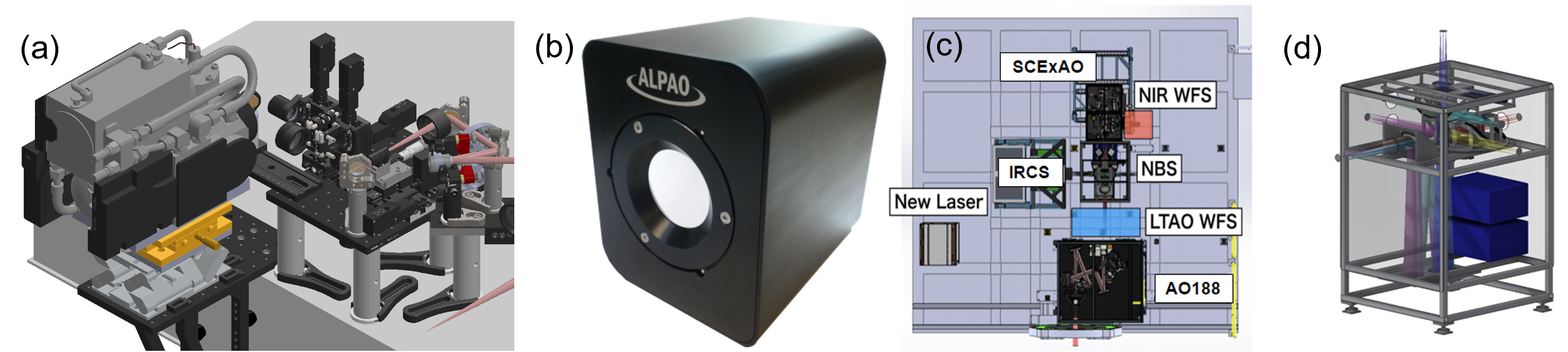}
    \vspace*{0.3cm}
    \caption{Hardware upgrades scheduled for the AO188: (a) NIR PyWFS, (b) new 64x64 ALPAO DM, (c) future configuration at the Nasmyth platform including a new beam switcher, allowing to split the light between the SCExAO and the IRCS, as well as other potential future instruments, and (d) a optomechanical design of the beam switcher.}
    \label{fig:new_upgrades}
\end{figure}

With the 8-meter telescope such as Subaru telescope, it is impossible to directly image Earth-like planets in the habitable zone of stars. However, the new generation of ELTs provides the necessary resolution to probe close to a significant number of M-type stars. Also, the development of high accuracy WFS\&C technologies is essential. One instrument that will target Earth-like planets in the habitable zone of M-type stars is the Planetary System Imager (PSI)\cite{fitzgerald2019planetary}, planned to be installed on the Thirty Meter Telescope (TMT)\cite{sanders2013thirty}. 
In order to get closer to the PSI, there are going to be some major upgrades on the Nasymth platform at the Subaru telescope in next few years\cite{ono2020overview}, as shown in Fig. \ref{fig:new_upgrades}. The first one is the addition inside the AO188 of a near-IR PyWFS that will benefit both the SCExAO and the InfraRed Camera and Spectrograph (IRCS)\cite{kobayashi2000ircs}.
The second one is the replacement of the current 188-actuator DM with the ALPAO 64x64 actuator DM. This will bring the number of actuators inside the pupil up to 3,000, and the new near IR PyWFS is also designed for this new DM, which means that when both of these modules are going to be installed inside the AO188, we would have an extreme AO capability right after the first stage of correction. Lastly, the third one is going to be the addition of a beam switcher on the Nasmyth platform at the Subaru telescope. This beam switcher will allow for a fast switching or even simultaneous between various instruments, such as SCExAO and IRCS, but also the addition of a new LTAO (Laser Tomography Adaptive Optics) WFSs and potential other visiting instruments.

In parallel to these major hardware changes, new advanced HCI technologies such as coherent differential imaging, predictive control, sensor fusion or real-time post-processing. In addition to demonstrate higher contrast in preparation for the PSI, the SCExAO will be able to directly image young Jupiter-mass planets closer to the habitable zone, down to ~3 AU, where they should be more abundant. This will give us more insight on the planet population around the habitable zone. Finally, a few older Jupiter-size planets should be reached by looking at the reflected light for the first time.

\section{Conclusion}
Over the last decade, SCExAO has been evolving into a unique platform, able to simultaneously perform competitive science using the classical HCI system, while testing new technologies and algorithms routinely for the HCI systems. This design allows for the team to collaborate with groups around the world on innovative technologies and algorithms. We are now testing new key technologies for future HCI systems for ELTs, such as MKIDs detectors, or fast and low-noise IR detectors. We also always welcome new algorithms, collaborations, and instruments. One goal of the SCExAO is to become a technology demonstrator and a testbed for the future PSI of the TMT. In order to achieve this, major upgrade changes in the SCExAO, but also in the AO188 and the whole Nasmyth IR platform of the Subaru telescope will modify the instrument configuration to get closer to the PSI for the TMT. Once the upgrades are completed, the combination of the SCExAO, the AO188, and the IRCS will be a fully complete system-level demonstrator for the future PSI of the TMT.

\acknowledgments 
 
This work is based on data collected at Subaru Telescope, which is operated by the National Astronomical Observatory of Japan.
The authors wish to recognize and acknowledge the very significant cultural role and reverence that the summit of Maunakea has always had within the Hawaiian community. We are most fortunate to have the opportunity to conduct observations from this mountain.
The authors also wish to acknowledge the critical importance of the current and recent Subaru Observatory daycrew, technicians, telescope operators, computer support, and office staff employees.  Their expertise, ingenuity, and dedication is indispensable to the continued successful operation of these observatories.
The development of SCExAO was supported by the Japan Society for the Promotion of Science (Grant-in-Aid for Research \#23340051, \#26220704, \#23103002, \#19H00703 \& \#19H00695), the Astrobiology Center of the National Institutes of Natural Sciences, Japan, the Mt Cuba Foundation and the director's contingency fund at Subaru Telescope.
KA acknowledges support from the Heising-Simons foundation. 

\bibliography{report} 
\bibliographystyle{spiebib} 

\end{document}